\documentclass[iop]{emulateapj}
\slugcomment{{\sc Accepted to ApJ:} August 31, 2012} 
\usepackage{graphicx,amsmath,amsfonts,amssymb,natbib,epsfig}

\newcommand{\kms}{\ km\ s$^{-1}$}

\def\squig{\sim\!\!}

\newcommand{\lt}{$<$}

\newcommand{\feh}{$\langle$[Fe/H]$\rangle$}
\def\Sec{${}^{\prime\prime}$\llap{.}}

\begin{document}

\title{Stellar Kinematics of the Andromeda II Dwarf Spheroidal Galaxy}

\author{Nhung\ Ho\altaffilmark{1},  Marla C.\ Geha\altaffilmark{1}}
\author{Ricardo\ R.\ Munoz \altaffilmark{2}}
\author{Puragra\ Guhathakurta\altaffilmark{3}}
\author{Jason S.\ Kalirai\altaffilmark{4,5}}
\author{Karoline\ M.\ Gilbert\altaffilmark{6,7}}
\author{Erik\ J.\ Tollerud \altaffilmark{1,7,8}}
\author{James S.\ Bullock \altaffilmark{8}}
\author{Rachael\ L.\ Beaton \altaffilmark{9}}
\author{Steve\ R.\ Majewski \altaffilmark{9}}

\altaffiltext{1}{Astronomy
  Department, Yale University, New Haven, CT 06520.
 ngocnhung.ho@yale.edu, marla.geha@yale.edu}
 
\altaffiltext{2}{Departamento de Astronom$\acute{\rm i}$a, Universidad de Chile, Casilla 36-D, Santiago, Chile}

\altaffiltext{3}{UCO-Lick Observatory, University of California,
   Santa Cruz, 1156 High Street, Santa Cruz, CA 95064.}

\altaffiltext{4}{Space Telescope
    Science Institute, 3700 San Martin Drive, Baltimore, MD 21218}
 
\altaffiltext{5}{Center for Astrophysical Sciences, Johns Hopkins University, Baltimore MD, 21218}

 \altaffiltext{6}{Department of Astronomy, University of Washington,
   Seattle, WA 98195}
   
  \altaffiltext{7}{Hubble Fellow}   
  
   \altaffiltext{8}{Center for Cosmology, Department of Physics and Astronomy, University of California, Irvine, CA 92697}

   \altaffiltext{9}{Department of Astronomy, University of Virginia,
   P.O. Box 3818, Charlottesville, VA 22903}

\begin{abstract}
\renewcommand{\thefootnote}{\fnsymbol{footnote}}

We present kinematical profiles and metallicity for the M31 dwarf spheroidal (dSph) satellite galaxy Andromeda II (And II) based on Keck DEIMOS spectroscopy of 531 red giant branch stars.  Our kinematical sample is among the largest for any M31 satellite and extends out to two effective radii ($r_{\rm eff} = 5.3'$ = 1.1\,kpc).  We find a mean systemic velocity of $-$192.4 $\pm$ 0.5 \kms\ and an average velocity dispersion of $\sigma_v = 7.8\pm1.1$\kms.  While the rotation velocity along the major axis of And II is nearly zero (\lt\,1\,\kms), the rotation along the minor axis is significant with a maximum rotational velocity of $v_{\rm max}=8.6 \pm 1.8$ \kms.   We find a kinematical major axis, with a maximum rotational velocity of $v_{\rm max}=10.9 \pm 2.4$ \kms, misaligned by 67$^\circ$ to the isophotal major axis.   And II is thus the first dwarf galaxy with evidence for nearly prolate rotation with a $v_{\rm max}/\sigma_v$ = 1.1, although given its ellipticity of $\epsilon$ = 0.10, this object may be triaxial.  We measured metallicities for a subsample of our data, finding a mean metallicity of [Fe/H] = $-1.39 \pm 0.03$\, dex and an internal metallicity dispersion of 0.72 $\pm$ 0.03 \,dex.  We find a radial metallicity gradient with metal-rich stars more centrally concentrated, but do not observe a significant difference in the dynamics of two metallicity populations.  And II is the only known dwarf galaxy to show minor axis rotation making it a unique system whose existence offers important clues on the processes responsible for the formation of dSphs.

\end{abstract}


\section{Introduction}

Dwarf galaxies are the most abundant galaxy sub-type in the Universe and, as a group, are incredibly diverse.  They range from dwarf irregulars (dIrrs) which are gas-rich, rotationally dominated systems to gas-poor, pressure-supported dwarf spheroidals (dSphs).  Similar to massive galaxies \citep{Dressler1980,ButcherOemler1984}, dwarfs also follow a morphology-density relation \citep{Ferguson91} with gas-poor, pressure supported systems preferentially crowding around a large parent galaxy while gas-rich, rotationally supported systems are further out from their host galaxy.  Within the Local Group several groups have shown \citep[e.g.,][]{Grebel2003, Grcevich2009} that the majority of dwarf galaxies near the Milky Way (MW) and the Andromeda Galaxy (M31) are gas-poor, while those further out tend to be more HI enriched. \citet{Geha2012} demonstrated that, within the Sloan Digital Sky Survey, dSph galaxies do not exist beyond a few virial radii from a massive host galaxy.  These studies suggest that environmental processes are transforming dIrrs into dSphs.
 
Dwarf galaxies are more susceptible to environmental effects as compared to their massive counterparts due to shallower potential wells.  The mechanisms responsible for this induced evolution depend on factors such as local density, orbital parameters, and the intrinsic properties of the dwarf.  Environmental effects from interactions with the host halo environment such as tidal stripping or tidal shocking may completely evolve a gas-rich dwarf into a dSph \citep{HodgeMichie1969,FaberLin1983, PiatekPryor1995,Gnedin1999,Munoz2008}.  Alternatively, a combination of tidal disruption and ram pressure stripping \citep{Mayer2006,Lokas2010,Kazantzidis2011} can work together to alter both the stellar kinematics and gas morphology of gas-rich dwarf galaxies.   

To study the effect of environment on dwarf galaxy evolution we look toward the Local Group, 
which hosts two massive systems, the MW and M31,  each hosting a rich array of dwarf satellites.   The MW's dSphs span the known range of dSph luminosities ($-1.5 > M_V > -18$), from ultra-faint galaxies with a few tens of established member stars \citep{Belokurov2006, Simon2007a} to bright systems such as the Large Magellanic Cloud.  Similar to the MW, M31 also possesses a diverse dwarf population spanning from low-luminosity dSphs ($M_V \sim -$6.5) such as And XXII \citep{Brasseur2011b} to M32, a compact elliptical ($M_V$= $-$16.5).  The similarities between the MW and M31 host system makes M31 a good complementary sample with which to supplement and expand our understanding of the properties and evolution the dwarf systems of more massive galaxies. 

Until recently, studies on the kinematics of the resolved stellar populations of the dwarf systems of M31 have been difficult due to their distance.   However, recent photometric and spectroscopic surveys of the M31 system have advanced our understanding of its dwarf satellites.  Current efforts include searching for new dwarfs and photometric characterization by the PAndAS survey \citep{Mcconnachie2009}, and spectroscopic studies of individual stars within these dwarfs by the Spectroscopic and Photometric Landscape of Andromeda's Stellar Halo (SPLASH) survey \citep{Kalirai2009,Kalirai2010,Tollerud2012}.  The success of these programs in both discovery and characterization now allows us to study in detail the morphological and kinematical properties of these dwarf galaxies.

One of the most luminous M31 dSph is Andromeda II (And II; $M_V = -$12.6).   At a distance of 650\,kpc from the MW, it is also one of the more nearby M31 satellites.  At first glance, And II is a typical dSph consisting of an older stellar population \citep{Dacosta2000} with little to no associated HI gas \citep{Grcevich2009,Lockman2012}.  However, deep photometric studies of its stellar populations have shown a stellar excess in the central regions of the surface brightness profile above a single exponential \citep{ Mcconnachie2006} along with a radial metallicity gradient with metal-rich stars more centrally concentrated \citep{Mcconnachie2007}.  These two observations point to evidence that And II may be comprised of two structural populations which have distinct spatial distributions and possibly distinct kinematics.  Initial kinematical studies of And II's resolved population by \citet{Cote1999b}, based on just seven stars, found a high velocity dispersion of $\sim$ 9 \kms\ along with a possible velocity gradient.  Later work by \citet{Kalirai2010} on And II's kinematics based on SPLASH spectroscopy of 95 members found a smaller dispersion of $\sim$7 \kms.  Despite the larger sample size, the spatial coverage of this dataset only sampled one quadrant of the And II surface and thus, provides an incomplete picture of the global kinematics of the system.

In this paper, we present the kinematical properties and spectroscopic metallicity of And II based on Keck/DEIMOS observations of 531 member stars, which includes stars in the SPLASH sample first presented by \citep{Kalirai2010}.  This represents the second largest kinematical dataset of any M31 dwarf satellite, just behind NGC 205 \citep{Geha2006a}.  The paper is organized as follows: \S\,\ref{methods} and \S\,\ref{sampsec} describe the spectroscopic and photometric reduction process, respectively.  \S\,\ref{sec_isophot_param} details our derivation of the position angle and ellipticity based on archival Subaru Suprime-cam data.  Using the previously derived position angle and ellipticity, we present the major and minor axis velocity profiles along with the global metallicity and chemodynamics in \S\,\ref{sec_results}. 

We adopt a distance modulus for And\,II determined by \citet{Mcconnachie2005a} via the tip of the Red Giant Branch (RGB) method of $(m -M)_0 =24.07 \pm 0.06$ ($652\pm 18$\,kpc).  This places And\,II at a distance of 185\,kpc from its parent galaxy M31.

\section{Spectroscopic Observations and Data Reduction} \label{methods}

\subsection{Target Selection} \label{targ_sec}

We select stars for spectroscopy using imaging data obtained in the Washington System \textit{M} and \textit{T$_2$} filters, and an intermediate-band DD051 filter, with the wide-field Mosaic camera on the Kitt Peak Mayall 4-meter telescope from 1998 to 2002 \citep{Ostheimer2003}.  More details can be found in a forthcoming paper (Beaton et al.~(in prep).  The DDO51 filter, centered near the surface gravity dependent Mgb and MgH absorption lines, allows us to separate foreground MW dwarf stars and target M31 giant stars \citep{Gilbert2006,Guhathakurta2006}.  Target selection and observing priority were based on a star's position on the Color Magnitude Diagram (CMD) relative to a metal-poor 13\,Gyr isochrone \citep{Girardi02} and position on the \textit{M$-$DDO51} vs.~\textit{M$-$T$_2$} color-color diagram \citep{Tollerud2012}.  

Due to the large chip gap in the DDO51 photometry which coincided with the central two arcminutes of And II, we augmented our photometry in these regions using Sloan Digital Sky Survey (SDSS) DR7 \citep{Abazajian2009} photometry in $g$ and $r$ bands.  This allowed us to spectroscopically sample the entire And II observable area, but with somewhat lower efficiency due the lack of ability to photometrically discern foreground dwarf stars from giant stars.  We use the position on the CMD and distance to a model isochrone to assign observing priority.  Out of the 12 spectroscopic masks described below, one was based on SDSS photometry.

  \subsection{Data Reduction} \label{sec_data}

The spectroscopic data were taken with the Keck~II 10-m telescope and the DEIMOS
spectrograph \citep{Faber2003a}.  Twelve multislit masks were observed
in And\,II between 2005 -- 2009.  Mask positions, exposure times and
other observing details are given in Table \ref{table_mask}.  The masks were observed
with the 1200~line~mm$^{-1}$\,grating covering a wavelength region
$6400-9100\mbox{\AA}$.  The spectral dispersion of this setup is
$0.33\mbox{\AA}$, and the resulting spectral resolution, taking into
account the anamorphic distortion, is $1.37\mbox{\AA}$ (FWHM,
equivalent to 47\kms\ at the Ca II triplet). The spatial scale is
$0$\Sec$12$~per pixel and slitlets were $0$\Sec$7$ wide.  The minimum slit
length was $4''$ which allows adequate sky subtraction; the minimum
spatial separation between slit ends was $0$\Sec$4$ (three pixels).

Spectra were reduced using a modified version of the {\tt spec2d} software pipeline (version~1.1.4) developed by the DEEP2 team at the University of California-Berkeley for that survey \citep{Davis2003,Cooper2012,Newman2012}. A detailed description of the two-dimensional reductions can be found in \citet{Simon2007a}.  The final one-dimensional spectra are rebinned into logarithmic wavelength bins with 15\,\kms\ per pixel.  Radial velocities were measured by cross-correlating the observed science spectra with a series of high signal-to-noise stellar templates. These stellar templates were observed with Keck/DEIMOS using the same setup as described in \citet{Geha2010} and cover a wide range of stellar types (F8 to M8 giants, subgiants and dwarf stars) and metallicities ([Fe/H] = $-2.12$ to $+0.11$).  We calculate and apply a telluric correction to each science spectrum by cross correlating a hot stellar template with the night sky absorption lines following the method in \citet{sohn2006a}.  We apply both a telluric and heliocentric correction to all velocities presented in this paper.

We determine the random component of our velocity errors using a Monte
Carlo bootstrap method.  Noise is added to each pixel in the
one-dimensional science spectrum, we then recalculate the velocity and
telluric correction for 1000 noise realizations.  Error bars are
defined as the square root of the variance in the recovered mean
velocity in the Monte Carlo simulations.  The systematic contribution
to the velocity error was determined by \citet{Simon2007a} to be
2.2\kms\ based on repeated independent measurements of individual
stars.  The systematic error contribution is expected to be constant
as the spectrograph setup and velocity cross-correlation routines are
identical.  We add the random and systematic errors in quadrature to
arrive at the final velocity error for each science measurement.

Radial velocities were successfully measured for 1613 of the 1643 extracted spectra across the twelve observed DEIMOS masks.  The majority of spectra for which we could not measure a redshift had insufficient signal-to-noise.  The fitted velocities were visually inspected to ensure reliability.  We include 84 stars with repeated spectra of which 8 showed velocity variations above expected errors.  We assume these systems are unresolved binaries and do not include them in the final sample.  For the remainder of the repeated sample, we take the velocity to be the weighted mean. Thus our final sample consists of 1566 unique velocities.

\section{Photometric Data Reduction} \label{sampsec}

We derive the ellipticity and position angle of And II for use in our kinematical analysis based on data collected at Subaru Observatory and obtained from the Subaru-Mitaka-Okayama-Kiso Archive (SMOKA), which is operated by the Astronomy Data Center, National Astronomical Observatory of Japan \citep{Baba2002}.  These observations were presented in \citet{Mcconnachie2007}, however they did not utilize the data for isophotal analysis.    These observations of And II cover an area of 0.5 deg$^2$ with 5 $\times$ 440 seconds in V and 20 $\times$ 240 seconds dithered exposures in $i'$.  

Preprocessing of the data were done by debiasing, trimming, flat fielding, and gain correcting each individual exposure chip-by-chip using median stacks of nightly sky flats.  The presence of scattered light due to bright stars both in and out of the field of view required us to remove this smoothly varying component before performing photometry and solving for a World Coordinate System solution (WCS).  To remove  scattered light, we fit the smoothly varying component by creating a $"\rm{flat}"$ for every chip within each frame by performing a running median with a box size of 300 pixels.  This was then subtracted from the original, unsmoothed frame to produce a final image for photometric processing.

Photometry was carried out using the DAOPHOT II package outlined in \citet{Stetson1993} using the method of \citet{Munoz2010}: first DAOPHOT/Allstar was run on all 25 individual object frames for V and $i'$ filters.  This produces a point spread function solution for each frame along with the associated starlist file containing x and y coordinates, magnitude and errors for all detections.  DAOMATCH was then used to group the 25 frames by their observing filters: five in V-band and 20 in $i'$-band.  DAOMASTER was run on the grouped frames to create two master star-lists, one for each filter, comprised of all objects which appear in two or more frames.  These two master lists were each then fed into ALLFRAME, which performs photometry on all input frames simultaneously and produces a finalized photometry file.  To collate the V and $i'$ photometry into one catalog, we run DAOMATCH to match up the filters and DAOMASTER to pick stars which appear in both master lists.  

We solve for the WCS solution by performing stellar matching between object frames and SDSS DR7 frames within the same region.  For every object reference frame, we detect at least 20 stars which appear in both the object frame and DR7 frames.  The third degree polynomial affine transformation between the two frames were calculated and iterated until the average scatter between the reference frame and transformed object frame was less than $0$\Sec$5$. The Subaru frames were then transformed into the DR7 frame and the associated DR7 astrometric solution was applied.  As a check, we ran DAOMATCH between the object and DR7 frames for the same stars and found the solutions to be very similar.  However, we were able to do this for only seven out of the 10 chip frames due to the lack of matching stars between frames.  We applied cuts to the final, transformed object catalog by keeping stars with DAOPHOT  $Chi$ and {\it sharp} values of $Chi \le$ 0.8 and $-0.5 \le$ \textit{sharp} $\le$ 0.5.

\section {Isophotal Parameters} \label{sec_isophot_param}

The major axis of And II has been determined previously in the literature by \citet{Mcconnachie2006} (hereafter, MI06).  In the sections below, we find a surprising result that the primary rotation axis is not along the major axis of And II.  To confirm that the major and minor axes are correctly determined, we re-determine these quantities using two methods: the IRAF task \textit{ellipse} in the STSDAS ISOPHOTE package \citep{Busko1996} and a circular annulus method.  We describe both methods below.

\subsection{IRAF \textit{ellipse}}

In our first method, we determine the surface brightness profile, ellipticity, and position angle as a function of radius using the IRAF task \textit{ellipse}.  Task \textit{ellipse} performs surface photometry using methods described in \citet{Jedrzejewski1987}.  2-D images are sampled along elliptical annuli to produce a 1-D intensity distribution as a function of position angle.  Best-fit ellipses are then determined by simultaneously fitting for the x and y centers, position angle, and ellipticity such that the intensity distribution is essentially flat, within a user-defined tolerance.  

\begin{figure}[!t]  
\centering
\includegraphics[width=.4\textwidth,angle=90]{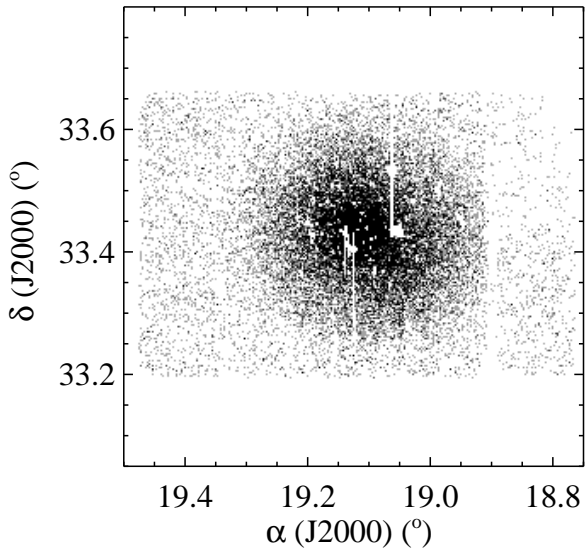}

\caption{Subaru Suprime-cam data binned into 10$''$ bins.  Saturated regions and those with incomplete coverage have been masked out (white).  The Subaru field completely covers the galaxy out to a radius of 15$'$, which is the edge of the detection area. }\label{fig_coverage}
\end{figure}

The package was designed for use on integrated photometry thus, we create an image by binning the data in $10''$ bins to produce a smooth light profile.  To account for gaps due to bright stars, we create a bad pixel mask to remove these regions from the ellipse fitting process.  The mask was created by selecting saturated regions in the image and expanding it by growing the region.  Large regions where detections were not possible, due to lack of sufficient spatial coverage, are also masked out by mapping their polynomial shape onto the mask.  The final binned image with aforementioned masked regions are shown in Figure \ref{fig_coverage}.  In total, roughly 6$\%$ of $\sim$ 30,000 stars in the sample were masked because they fell into the same bins as the masked regions.  While non-negligible, the percentage is low enough such that the statistics of the fit would not be greatly affected.

\begin{figure}[!t]
\centering
\includegraphics[width=.33\textwidth,angle=90]{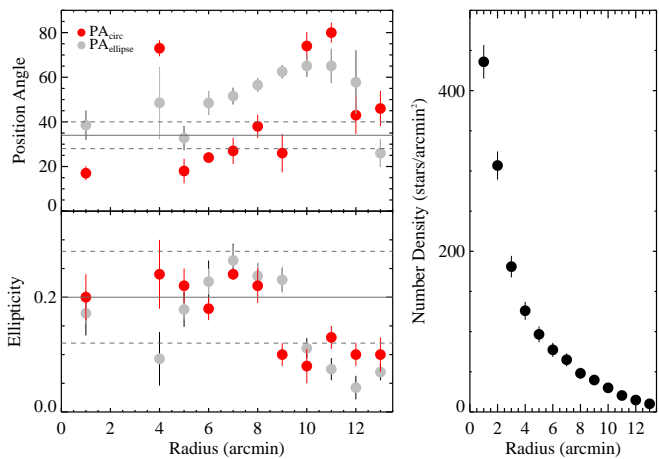}
\caption{Surface brightness profile ({\it right}), Position Angle ({\it left, top}) and ellipticity ({\it left, bottom})  as a function of radius for And II using IRAF \text{ellipse} (grey circles) and the circular binning method (red circles).  Solid grey line denotes the PA value derived by MI06 with the associated error represented by dashed grey lines.  }\label{pavsrad}
\end{figure}

\begin{figure}[!h]
\centering
\includegraphics[width=.4\textwidth]{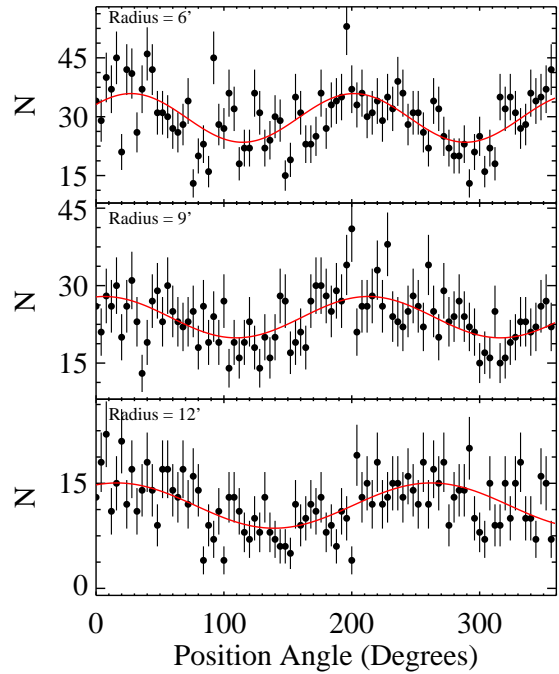}
\caption{ Intensity versus Position Angle for three sample annuli at 6$'$, 9$'$, and 12$'$.  Data are shown in filled, black circles with associated poisson errors.  The red curves represent the best-fit,  general sine curves to the observed distribution.  The resulting fitted PA values for annuli at 6$'$, 9$'$, and 12$'$ are 24$^\circ$, 26$^\circ$, and 43$^\circ$, respectively. }\label{circ_ann}
\end{figure}

The x and y center were independently derived by collapsing the data in the RA direction and the DEC direction and binning in 10$"$ bins, which was chosen to match the resolution of the input image for task \textit{ellipse}.  We then performed a non-linear least squares fit to the intensity distribution along the RA and DEC directions using a Lorentzian model profile.  To account for the effects of the previously mentioned data holes, we weight the fit to account for these low-count regions.  The derived center is RA = 19.1089$^\circ$$\pm$ 6.92$''$ and DEC = 33.4270$^\circ$$\pm$6.9$''$.  Using fixed coordinates for the center and a step size of 60 arcseconds, we derive the position angle as a function of radius for each annulus until the edge of the detection area is reached.  This step size was chosen because it is wide enough to encompass the large data gaps in the central regions due to bright, foreground stars, but still allow for enough data pixels to properly characterize the shape within the annular region.  

We find a position angle of 26$\pm$6$^{\circ}$ at the outer radius of 13$'$; this is consistent, within errors, to the value of 34$\pm$6$^{\circ}$ found by MI06, who only derived isophotal parameters for the outermost isophotes .  We do not include the PA and ellipticity at 2$'$ and 3$'$ due to large uncertainties, which were a result of extensive masking of the region because of bright, foreground stars.  Task \textit{ellipse} also allows for the calculation of the ellipticity, $\epsilon$ = 0.07$\pm$0.015; less than that found by MI06 in which $\epsilon$ = 0.20$\pm$0.08.  Figure \ref{pavsrad}, \textit{left} shows the position angle and ellipticity as a function of radius derived from \textit{ellipse} represented by filled, grey circles.  Based on our results, And II is more round at the outer isophotes than was previously observed.  This difference may be due to the coverage in MI06 not encompassing all of the And II field, as compared to the Subaru observations used in this work. 

\subsection{Circular Annulus} \label{circ_ann_der}

The circular annulus method is similar to \textit{ellipse} in that the isophotal position angle and ellipticity for each annulus is derived from fits to the distribution of intensity as a function of position angle.  However, while \textit{ellipse} takes as input an image, the circular annulus method is performed using individual stars.  Thus, the circular annulus method is better tuned to low-surface brightness galaxies, such as dSphs, where a smooth light profile may not be present.  Additionally, while \textit{ellipse} adjusts the shape of the sampling annulus as it converges to a solution for each step in radius, this method samples the distribution in concentric, circular annuli from a fixed center.   Using a series of model galaxies derived from the surface brightness profile of And II, we determined the position angle and ellipticity by comparing the observed And II intensity versus position angle distribution to the modeled distribution, which is described in detail in the following paragraphs.  

From the fixed center, obtained using the previously mentioned Lorentzian fit to the 2-D light profiles, we increase in 1$'$ annular steps until the edge of the detection area is reached.  Thus, the first annulus encompasses the circular region within 1$'$.  The data within this annulus is then binned into one-degree position angle bins resulting in an intensity versus position angle distribution.  To simultaneously fit the position angle and ellipticity for each annulus, we create model galaxies with the same surface brightness profile as And II (Figure \ref{pavsrad}, \textit{right}), but with ellipticities ranging from 0.0 to 0.5 in steps of 0.01 and position angles which vary from 0$^{\circ}$ to 180$^{\circ}$ in steps of 1$^{\circ}$.  To quantify the goodness of fit, each annular intensity versus position angle profile was fitted to a general sine curve of the form $$ F(x) = A + B\times \sin(C\times(x-\phi))$$ where A is the baseline, B is the amplitude, C is the period of the curve, and $\phi$ is the phase shift from a standard, unshifted sine wave.  Figure \ref{circ_ann} shows a sample of three position angle versus intensity profile and the best-fitting sine curve for the observed And II distribution.  The model intensity versus position angle profiles at each annulus are then compared to the observed And II profiles using the three free-parameters of the general sine curve: the amplitude B, the period C, and the phase shift $\phi$.  The resulting best-fit model profile have parameter values which are within a certain tolerance of the best-fit parameters for the observed profile.

Figure \ref{pavsrad} shows the position angle as a function of radius of And II using the circular annulus method (red circles).  The PA and ellipticity at the outermost radius of 13$'$ derived from this method are 46$^\circ$$\pm$8$^{\circ}$ and 0.10$\pm$0.02 and are consistent, within errors, to the published value.  While it is on the high side, we will show in later sections that this does not greatly affect the observed dynamics of the system.  This method serves as an independent method to verify the position angle and ellipticity outputted by \textit{ellipse}.  Table \ref{isophotal_parameters} lists the results from both methods along with the number density profile.  For the remainder of this paper we adopt a value of PA =  46$^\circ$$\pm$8$^{\circ}$ and ellipticity= 0.10$\pm$0.02 for And II.  

\begin{figure}[!h]
\centering
\includegraphics[width=.35\textwidth,angle=90]{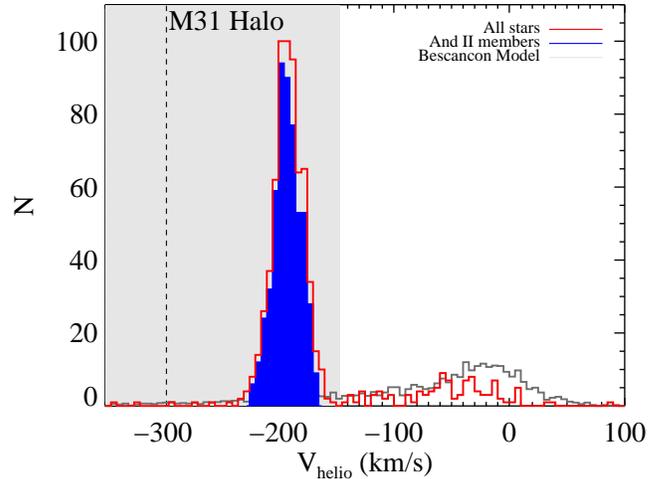}

\caption{Heliocentric velocity histogram showing all stars with a measured redshift (red).  The filled (blue) histogram shows the distribution for our final sample of And II members.  The Besan\c con model distribution of MW foreground stars in shown in grey.  Also shown in shaded grey is the region encompassed by the M31 halo with dashed-line representing the center of the distribution.}\label{member_select}
\end{figure}

	
\begin{figure*}[!b]  
\centering
\includegraphics[width=.7\textwidth]{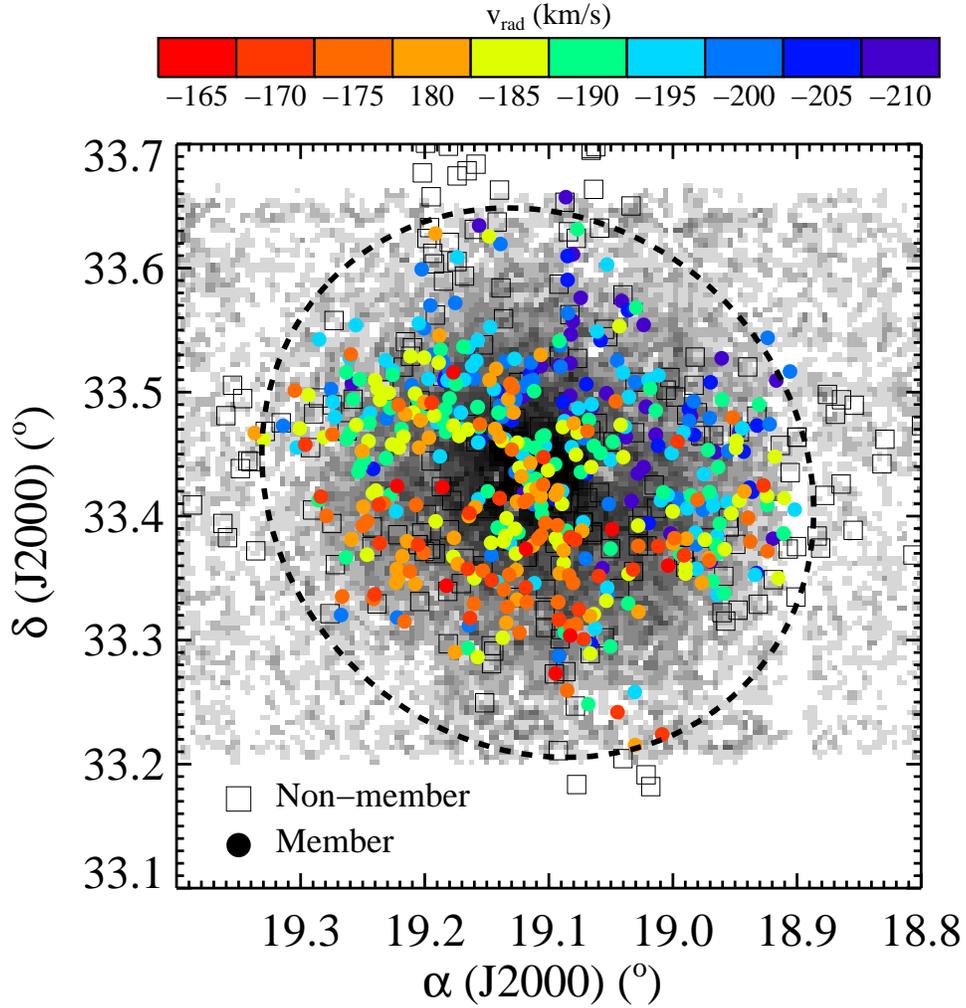}

\caption{Spatial position of observed stars overlaid on a binned Subaru image with dashed ellipse corresponding to 2.5$r_{\rm eff}$.  Black squares correspond to stars identified as non-members while filled circles correspond to members, color coded by velocity.} \label{fig2}
\end{figure*}

\section {Spectroscopic Results} \label{sec_results}

\subsection{Membership Selection} \label{mem_selection}

\begin{figure*}[!t]
\centering
\begin{tabular}{ccc}
\epsfig{file=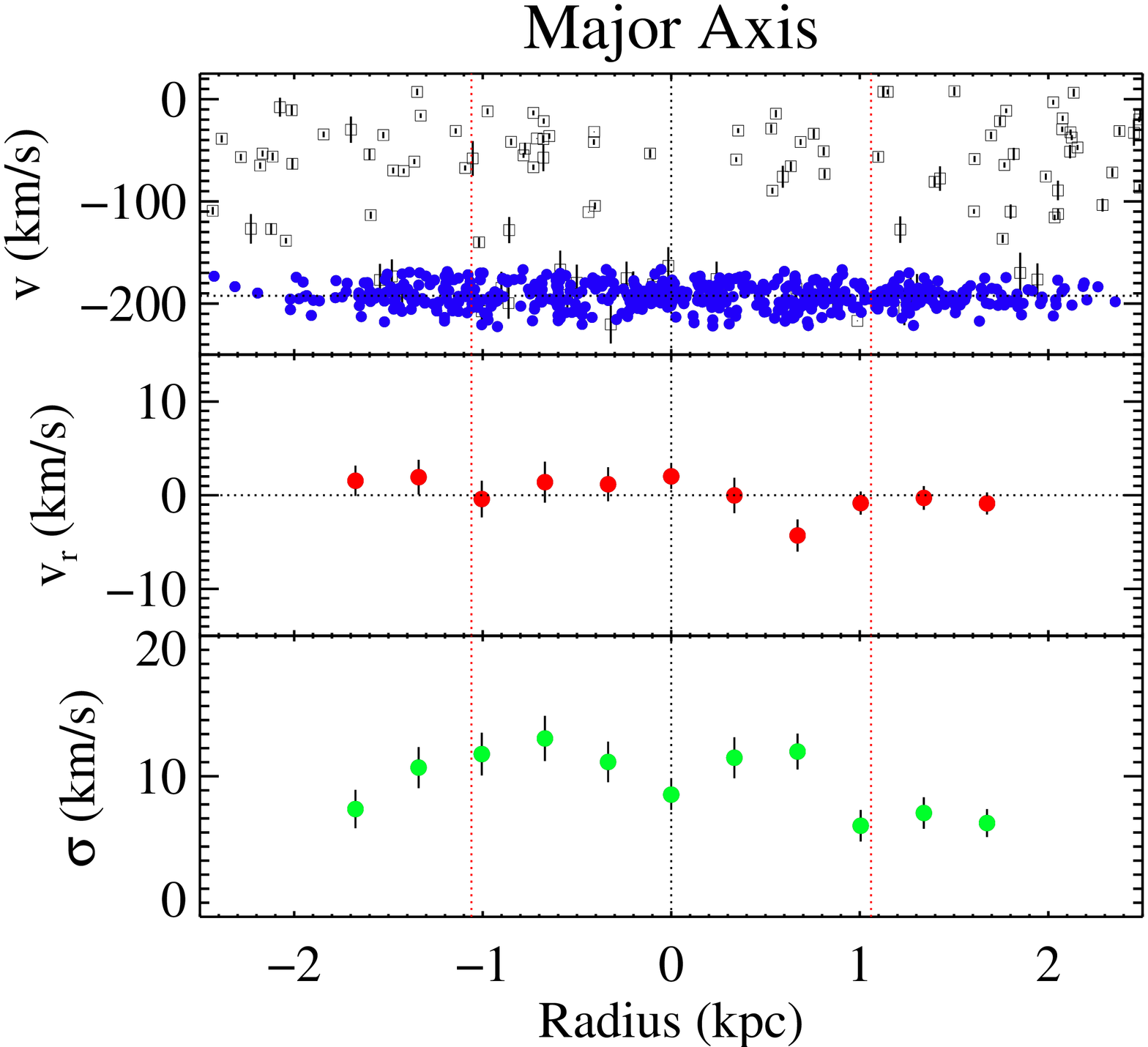,width=.4\textwidth,angle=90}
\epsfig{file=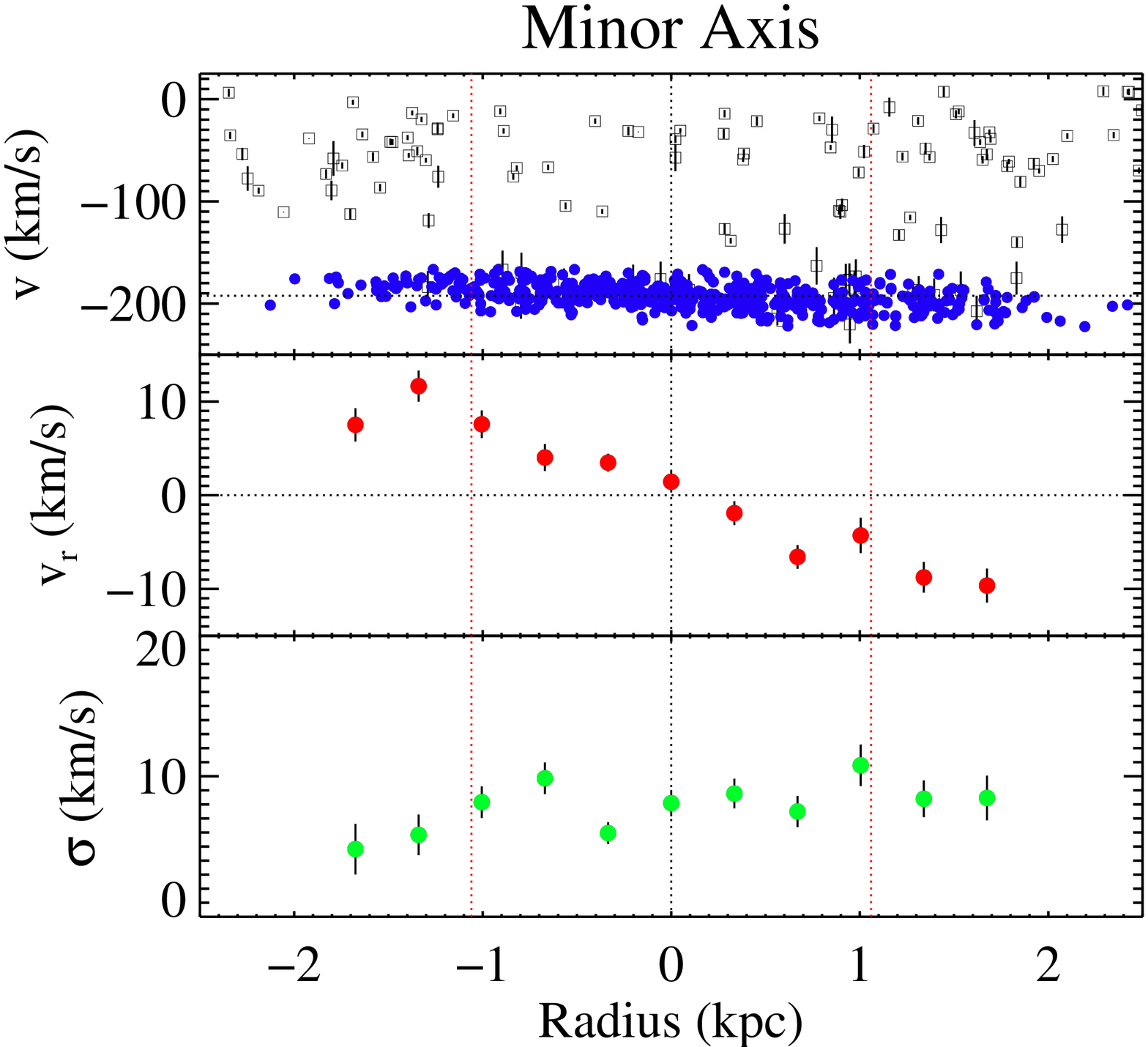,width=.4\textwidth,angle=90}
\end{tabular}
\caption{Major and Minor axis profiles for And II.  Top: individual velocities as a function of semi-major axis distance for member stars (filled blue circles) and non-members (open squares).  Middle: binned rotational velocity as a function of radius showing possible weak residual rotation in the central regions with no observed rotation at large radii.  Bottom: binned dispersion profile consistent with a flat profile out to last radial bin.  Dot-dashed red lines represent the effective radius, derived in MI06.} \label{axial_profiles}
\end{figure*}

Measuring velocities of individual stars allow us to probe the internal kinematics of dSphs to a much larger radius than possible using integrated light spectroscopy.  To properly characterize the kinematics, membership determination of these individual stars is critical.  The velocity distribution of And II overlaps with the wings of the stellar velocity distribution of both M31 and the MW (Figure~\ref{member_select}), thus a simple cut in velocity alone is not enough to determine membership.  To establish membership we use cuts based on three priors which are described in more detail in \citet{Gilbert2006}: (i) star's position in the Color-Magnitude Diagram, (ii) line of sight velocity, and (iii) the strength of the Na I absorption line at $\lambda$8190 \AA, which is a dwarf-giant discriminator with foreground dwarfs possessing a higher Na I equivalent width than RGBs. 

To determine velocity membership cuts, the heliocentric velocity distribution of the two major contamination sources, M31 and the MW, were examined.  The M31 halo peaks at a heliocentric velocity of $-297$ \kms\ \citep{devaucouleurs1991} with a broad one $\sigma$ spread of 150$^{+50}_{-30}$ \kms\ \citep{Reitzel2002}.  Despite the broadness of the M31 distribution, the expected number of M31 RGB contaminants should be low, about 1 star per DEIMOS mask \citep{Gilbert2006}, because of And II is in the outskirts of the observed M31 halo at a projected distance of 185\,kpc. Foreground contamination from Milky Way disk, spheroid, and halo stars should also be low at the position of And II.  The Besan\c con MW model \citep{Robin2003} at And II's Galactic position peaks at a heliocentric velocity of $-$7\,\kms\ with a dispersion of 29\,\kms.  And II's velocity distribution peaks at $-$192.4 \kms, which is outside the 5-$\sigma$ wing of the expected MW velocity distribution.  The percentage of MW stars which have velocities within 3.5-$\sigma$ of the peak of the And II distribution is 1.7$\%$ of the total MW distribution, thus we expect no more than a few foreground velocity contaminants.  From this, we then define stars with radial velocities within 3.5-$\sigma$ of the peak And II velocity distribution, $-$228 $< v <$ $-$157 \kms, as possible members.  While a simple kinematical selection is able to remove a majority of MW and M31 contaminants, there is a small fraction that may still be present in the culled sample.  Comparing the strength of the Na I line as a function of color, we use a simple cut of V-I $<$2.5 and EW$_{Na I} <$ 4 which further eliminates an additional 14 likely MW foreground stars from the sample; we note this number is consistent with that predicted by the Besan\c con model.  Figure \ref{member_select} shows the velocity histogram of the final And II sample, which consists of 531 And II member RGB stars, along with the M31 halo distribution width and MW Besan\c con model.

\subsection{Internal Kinematics}

The data coverage of our spectroscopic members extends out to 15$'$, corresponding to a radius of $~$2.5$r_{\rm eff}$ or 2.7 kpc, and spatially covers the observable surface of And II (Figure~\ref{fig2}).  We calculate the systemic velocity of And II using the maximum likelihood method described in \citet{Walker2006} which assumes that the observed dispersion is the sum of the true intrinsic dispersion and the velocity errors observed.  Using this method, we find a systemic velocity of $v_{\rm sys} = -192.4 \pm 0.5$ \kms.  The derived $v_{\rm sys}$ is consistent, within errors, to previous studies which find a value of $v_{\rm sys} = -188.5 \pm 3.6$ \kms\ \citep{Cote1999b} and $v_{\rm sys} = -193.6 \pm 1.0$ \kms\ \citep{Kalirai2010}, the latter of which is a subsample of our data.

Color coding member stars by their radial velocities in Figure~\ref{fig2}, we see clear evidence of a rotational signal extending out to $\squig$ 2.5$r_{\rm eff}$.  To derive the velocity and dispersion profiles for And II, we bin the individual velocity measurements based on axial distance, \textit{x}, using bin widths $\Delta x$=0.25$r_{\rm eff}$.  The bin with number \textit{i} spans values of x from $(i-\frac{1}{2})\Delta x$ to $(i+\frac{1}{2})\Delta x$.  The value of \textit{x} for each star was calculated by compressing the stars along the desired axis and using its distance to the center as its axial distance.  We place an additional requirement that each bin contain at least 25 stars to ensure that the derived velocity and dispersion are robust.  The mean velocity and velocity dispersion within each bin were determined using the previously mentioned maximum likelihood method.  In the following sections we present velocity profiles for And II along with a discussion.

\subsubsection{Velocity Profiles} \label{velo_profiles}

The rotational velocity and velocity dispersion as a function of radius along the major axis are shown in the left panel of Figure \ref{axial_profiles}, \textit{left} and Table \ref{major_axis_data}.  The rotation curve shows a small amount of rotation in the central regions and no rotational signal at the outer radii.  The velocity dispersion appears relatively flat with a slight decrease at the outer radii.  Folding the profiles along their symmetry axis using a weighted sum, we derive a maximum rotational velocity of $v_{\rm max}$ = 0.78$\pm $1.23 \kms.  Using a weighted mean, we find an average velocity dispersion of $\langle \sigma_v \rangle$ = 8.97$\pm$1.24 \kms.  While this result would be consistent with kinematical profiles of other Local Group dSphs with similar luminosity \citep{Walker2006,Battaglia2006}, it is clear from Figure \ref{fig2} that this is not a good description of And II's kinematics

We next derive the velocity and dispersion profiles along the minor axis of And II.  The minor axis profile, seen in the right panel of Figure \ref{axial_profiles}, \textit{right} and Table \ref{minor_axis_data}, shows a strong velocity gradient with a maximum rotational velocity, $v_{\rm max}$ = 8.6$\pm$1.8 \kms\ and a flat velocity dispersion with an average dispersion, $\langle \sigma_v \rangle$ = 7.8$\pm$1.1 \kms.  The average dispersion is consistent with previously reported values of $\sigma_v = 7.1 \pm 2.1$ \kms\ \citep{Cote1999b} and $\sigma_v = 7.3 \pm 0.8$ \kms\ \citep{Kalirai2010}.  The \citet{Cote1999b} study contained only a handful of member stars and the \citet{Kalirai2010} sample contained only the first quadrant of the And II.  Thus, we report, for the first time, strong minor axis rotation in And II.   Since we do not observe a flattening of the rotation curve at the outer radii or a turnover, our derived value for $v_{\rm max}$ is likely a lower-bound to the true maximum rotational velocity.

\begin{figure}[!t]
\centering
\includegraphics[width=.35\textwidth,angle=90]{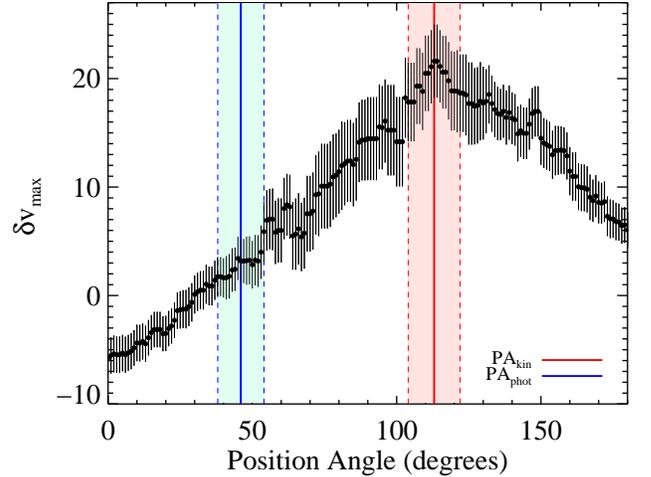}

\caption{The rotation velocity gradient, $\delta v_{\rm max}$ as a function of position angle for the observed And II spectroscopic velocity sample.  The kinematical major axis PA is denoted by the solid red line with error bounds denoted by dashed red lines.  The photometric major axis derived using the circular annulus method is denoted by the solid blue line with error bounds denoted by dashed blue lines.} \label{fig_kinematic_axes}

\end{figure}

And II's strong rotation along the minor axis is indicative of a misalignment between the kinematical and photometric major axes.  To determine the extent of this misalignment, denoted as $\Psi = |\rm{PA}_{\rm kin}-\rm{PA}_{\rm phot}|$, we determine the position angle of the kinematical major axis.  Assuming that the kinematical position angle is coincident with the position angle where the velocity gradient is largest, we produced velocity profiles for all position angles between 0$^\circ$ $<$ PA $<$ 180$^{\circ}$ with increments of 1$^\circ$.  Using the factor, $\delta v_{\rm max}$, to quantify the strength of the velocity gradient from the northeast side to the southwest side, we determine the position angle where $\delta v_{\rm max}$ is maximized.  The error associated with this  was determined using a Monte Carlo bootstrap method.  By adding to the binned velocity a random number which falls within the bounds of the derived $\pm$1-$\sigma$ errors from the maximum likelihood analysis, we build up a distribution of position angles where $\delta v_{\rm max}$ is maximized for 1000 iterations.  The square root of the variance of this distribution gives our formal 1-$\sigma$ errors on the position angle of the kinematical axis.  We find a kinematical position angle PA$_{\rm kinematic}$ =  113$\pm$9$^{\circ}$ with a maximum rotational velocity of  $v_{\rm max}$ = 10.9$\pm$ 2.4 \kms\ .  This is very inclined with the previously derived photometric major axis (\S\,\ref{circ_ann_der}) .  We show graphically this misalignment in Figure \ref{fig_kinematic_axes} which plots $\delta v_{\rm max}$ as a function of position angle with the derived kinematical major axis and its associated errors in red and the derived photometric major axis and its associated errors in blue.  Quantifying this, the degree of misalignment between the photometric major axis and kinematical major axis is $\Psi = 67^\circ \pm 12^\circ$.

\subsection{Prolate Rotation in And II}

We have shown that the kinematical major axis of And II is nearly coincident with the photometric minor axis.  In other words And II exhibits prolate rotation.  This is the first case of prolate rotation observed in a low-luminosity system, although cases of minor axis rotation have been well documented among elliptical galaxies.  \citet{Franx1989} showed that the percentage of elliptical galaxies with kinematical axes which align with the photometric minor axis is higher than expected for systems which are all oblate rotators.  \citet{Capellari2007} showed that systems where $\Psi$ is large are also systems which have low ($v/\sigma)$ and typically low ellipticity ($\epsilon < 0.3$) and fall at or below the predicted relation for isotropic rotators.  Further quantifying this with a larger sample, \citet{Krajnovic2011} showed that in the ATLAS$^{3D}$ sample of 260 early-type galaxies, only 10$\%$ of galaxies in their observed sample had $\Psi > 15^{\circ}$.  Of those 26 galaxies, only 9 have $\Psi > 65^{\circ}$, out of which only one, NGC 5485, can be classified as a rotationally dominated system and all with $\epsilon < 0.3$, similar to the observed ellipticity of And II.  The low numbers of such misaligned systems in more massive and more well studied galaxies point to the rarity of these systems, with systems which exhibit fast, minor-axis rotation even rarer.  Within the Local Group, the small handful of satellites which show rotation all have aligned kinematical and photometric major axes.  And II's clear misalignment between its kinematical and photometric major axes along with its quite strong rotation makes it a clear outlier.

\begin{figure}[!h] 
\centering
\includegraphics[width=.35\textwidth,angle=90]{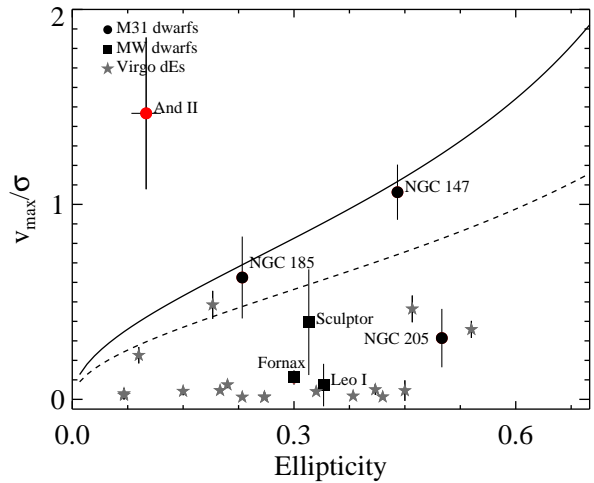}

\caption{Ratio of maximum rotational velocity divided by mean velocity dispersion vs ellipticity for M31 dwarfs (filled circles), MW dwarfs (filled squares) and Virgo dEs (filled stars).  The solid line represents the relation for a rotationally supported isotropic oblate spheroid while the dashed line represents the relation for a rotationally supported isotropic prolate spheroid.}\label{fig_vsigma}
\end{figure}

We compare And II's observed dynamics to Milky Way dSphs which are at similar host distances and luminosities: Sculptor, Fornax, and Leo I \citep{Walker2006}, along with its M31 neighbors \citep{Geha2010} and Virgo dEs \citep{Geha2003}.   Figure \ref{fig_vsigma} shows the relation between ellipticity and maximum rotational velocity divided by velocity dispersion.  The solid line denotes the relation between $(v_{\rm max}/\sigma)$ and ellipticity for a rotationally supported oblate spheroid while the dashed line is the relation for a rotationally supported isotropic prolate spheroid \citep{Binney1978} .  NGC 147 and NGC 185 are slightly more luminous M31 satellites as compared to And II and show rotational support on their major axis, consistent with an oblate sperhoid.  If we compare And II to its Milky Way counterparts Leo I and Fornax, it is clear that these MW satellites show a lack of rotational support and are well below the predicted relation for both prolate and oblate rotationally supported spheroids.  Sculptor, a MW satellite which has been shown to possibly possess a velocity gradient \citep{Battaglia2008b}, is still well within what has been observed for other dwarfs.  And II, however, lies above predicted relations for both prolate and oblate spheroids.  Taking the ratio between the observed $(v_{\rm max}/\sigma)$ = 1.4 for the kinematical major axis, and the predicted $(v/\sigma_{\rm iso})$, the resulting $(v/\sigma)_{\star}$ = 5.3 for an isotropic prolate spheroid, the resulting relation shows a quite large mismatch.  Given that the plotted relation is for isotropically supported objects, any anisotropy in the system would further flatten the relation. 

In the previous paragraphs we discuss the observed prolate rotation, but do not attempt to constrain whether the rotation is that of an axisymmetric, prolate rotator or a triaxial galaxy.  If And II is an axisymmetric, prolate rotator, then the rotation observed is intrinsic to the system and the observed major and minor axes represent the true major and minor axes of the system.  However, if it is triaxial there are two possible explanations: one intrinsic and one due to projection effects.  In a triaxial system, the angular momentum vector can lie anywhere in the plane of the short and long axes, thus an intrinsic misalignment between the rotational and short axis is possible.  In projection, a triaxial galaxy almost always has an apparent minor axis that is at a different position angle to the projected short axis of the galaxy \citep{Franx1991,Binney1985}.  Thus, if the galaxy rotates about its short axis, the observer will measure a gradient along the apparent minor axis.  To determine the shape of And II, more detailed kinematical modeling is required.  

 \subsection{Metallicity} \label{metallicity_sec}

We measured the mean metallicity and chemo-dynamics for a 477 member subset of our sample using the equivalent width (EW) of the Ca II lines.  

\subsubsection{Metallicity Calibration}
The  method outlined by \citet{Rutledge1997} has been the standard used to calculate metallicity using the EWs of the Ca II lines.  While the Rutledge calibration works well for most stars, it fails at low metallicities where the Ca II lines are much broader and non-LTE effects begin to dominate in the stellar photosphere.  In our analysis, we use an updated empirical calibration from \citet{Starkenburg2010}, which produces low metallicity stars ([Fe/H]$<-$2) while still being consistent with the Rutledge calibration at the high metallicity end.  This calibration has been shown to be robust in many different systems \citep [e.g.][]{Battaglia2011,Tafelmeyer2010}

\subsubsection{Global Metallicity Properties}

Using the Starkenburg CaT calibration, we measured a mean metallicity for And II of \feh  = $-1.39\pm $0.03\,dex and metallicity dispersion of $\sigma_{\rm{[Fe/H]}}$ = 0.72$\pm$ 0.03\,dex using the maximum likelihood analysis similar to that employed in our velocity calculations.  Our values for \feh  are consistent with previous photometric estimates:\feh = $-$1.59$ ^{+0.44}_{-0.12}$ \citep{Konig1993}, \feh = $-$1.49 $\pm$ 0.11 \citep{Dacosta2000}, and \feh = $-$1.5 $\pm$ 0.1 \citep{Mcconnachie2007}.  The measured mean metallicity is also consistent with published spectroscopic metallicity calculations: \feh =  $-$1.47 $\pm$ 0.19 \citep{Cote1999a} and \feh = $-$1.64 $\pm$ 0.04 \citep{Kalirai2010}, which is a subsample of our full dataset.  While our mean metallicity is consistent with previously derived values, our calculated metallicity dispersion is larger by $\squig$ 0.35\,dex than previous photometric and spectroscopic work.  This increase in metallicity dispersion is not surprising given our larger sample size, which includes more metal-rich stars than previous works,  and the nature of the Starkenburg calibration, which extends the classic calibration relation down to more metal poor stars.  Given the increase in the number of stars which inhabit the metal poor tail of the metallicity distribution function, an increase in the metallicity dispersion is expected.

\subsubsection{Chemo-Dynamics}

\begin{figure}[!t]
\centering
\includegraphics[width=.35\textwidth,angle=90]{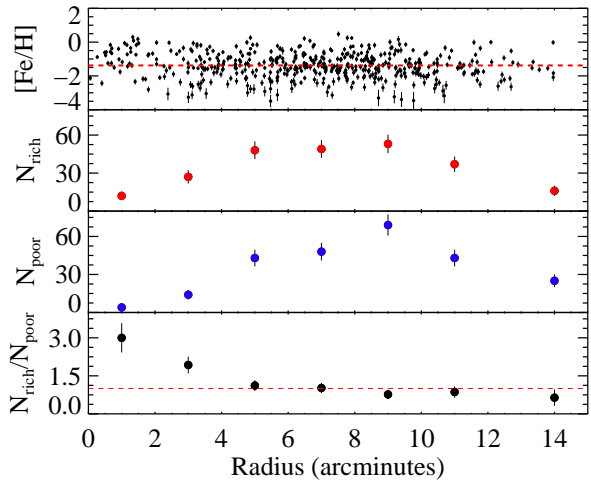}

\caption{Number of stars within elliptical annuli for both metal rich and metal poor populations. Top: Metallicity of individual stars as a function of elliptical distance.  Middle, top: number density of metal-rich stars as a function of elliptical distance.  Middle, bottom: number density of metal-poor stars as a function of elliptical distance.  Bottom: Ratio of metal rich to metal poor stars as a function of elliptical distance.  We see a radial metallicity gradient with more metal rich stars residing in the central regions of And II than metal poor stars.} \label{chemo_dyn}
\end{figure}
In the Local Group, radial metallicity gradients have been observed both photometrically and spectroscopically in many dSph such as Sculptor \citep{Battaglia2006}, Sextans \citep{Battaglia2011}, and Fornax \citep{Tolstoy2004}.  The presence or absence of a metallicity gradient, or lack therof, are clues to the star formation, chemical enrichment, and dynamical history of these objects because metallicity is encoded into stars at their formation.  To explore whether there is a presence of a radial metallicity gradient in our sample, we split our sample into two components: a metal rich component which has [Fe/H] $> - $1.39 and a metal poor component with [Fe/H] $< - $1.39.  From the fitted center, the number of metal rich and metal poor RGB stars were counted in elliptical annuli of width = 1$'$.  We find that the number of metal rich stars outnumber metal poor stars in the innermost two arcminutes by a factor of $\squig$ 2:1, as shown in Figure \ref{chemo_dyn}.  This overdensity of metal rich stars relative to metal poor stars persist out to five arcminutes, at which point the number of both populations appear to be evenly distributed.  Previous photometric work by M07 on the radial distribution of metal rich versus metal poor stars also found similar results with metal rich stars more centrally concentrated than metal poor.  The origin of this population gradient can give us clues into the star formation history of the galaxy.

\begin{figure}[!t]
\centering
\includegraphics[width=.35\textwidth,angle=90]{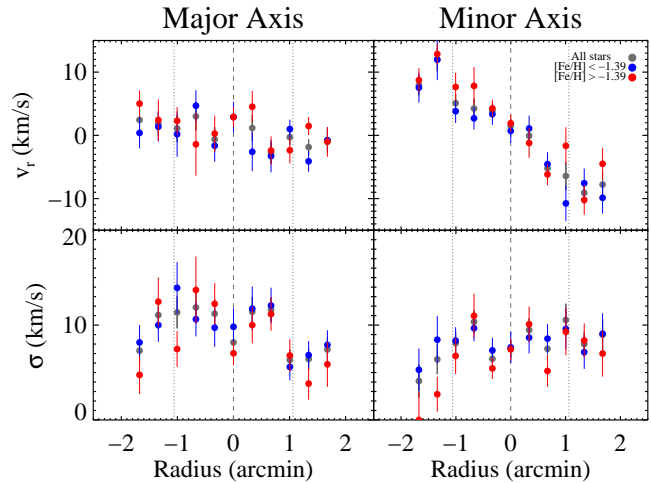}

\caption{Stellar kinematics of And II, split into two populations.  The population which has metallicity higher than the mean is shown in red while that which has metallicity lower than the mean is shown in blue.  For both major and minor axis profiles, there is no observed difference in the dynamics of these two metallicity populations.}\label{radial_metals}
\end{figure}

M07 postulated that the presence of a radial metallicity gradient between populations of different ages was due to the presence of two distinct stellar populations which had distinct dynamics.  Taking metallicity as a proxy for age, we use our RGB sample to determine whether And II's different RGB populations are kinematically distinct.  To examine this, we again split our sample into metal rich and metal poor by the mean metallicity.  The kinematical analysis detailed previously are repeated on these two sub-samples and velocity and dispersion profiles were produced for both major and minor axes, shown in Figure \ref{radial_metals}.  We find no statistically significant difference in the kinematical behavior of the metal rich and metal poor components along both axes with both populations exhibiting strong rotation along the minor axis and no rotation along the major axis.  Thus, while And II does exhibit a radial metallicity gradient, there is no evidence that the populations which gave rise to this gradient possess different kinematics.  This is in contrast to studies by \citet{Battaglia2006} on Fornax, \citet{Tolstoy2004} on Sculptor, and \citet{Battaglia2011} on Sextans which found that populations which exhibit radial metallicity gradients also exhibited distinct kinematics.  However, similar studies of Leo II \citep{Koch2007a,Koch2007c}, and Leo I \citep{Koch2007b}, which exhibit mild to no radial metallicity gradient, showed no differences between the metal rich and metal poor populations.

\section{Dynamical Mass}

We have demonstrated that And II is rotationally supported with a well behaved rotational profile along the minor axis.   In M31,  slightly brighter satellites than And II show clear evidence of rotation (NGC 205, NGC 147, NGC 185 and M32), while the fainter satellites around both M31 and the MW are dispersion supported.   For both these types of systems, it is possible to determine the dynamical mass assuming dynamical equilibrium and the Jeans Equation \citep{Walker2009, Geha2010, Wolf2010}.    However, determining the mass for And II is complicated by the presence of rotation along the minor axis.     The lack of observed prolate systems which have $(v/\sigma)_\star > 1$ may mean that the lifetime of systems with such a configuration is quite short.   Without full dynamical modeling, outside the scope of this paper, we can not determine the dynamical stability of such a system.  However, attempting to place And II into context with other dSphs, we present below mass estimates based on two simple mass estimators

If we assume that And II is in dynamical equilibrium, we can determine the mass using two mass estimators.  The first method is to assume that the observed line of sight velocity and velocity dispersion represent the total rotation velocity and velocity dispersion.  Under this assumption, we can include the increased dynamical support from the dispersion component into the velocity budget such that $v_{c} = \sqrt{v_{los}^{2}+\sigma_{los}^2}$.  The enclosed mass of the system is described as M($<$r) = $\frac{v_c^2r}{G}$.  We calculate the mass at the half-light radius of $r_{\rm eff} = $ 1.06 kpc.  Taking into account the effect of inclination, we deproject the half light radius using the observed axial ratio.  Using a minimum disk ratio $q_{0}$ = 0.22 \citep{devaucouleurs1991}, the inclination angle for our system is $\it{i} = \rm 40^{\circ}$.  Deprojecting the half light radius using this inclination, we get  $r_{\rm eff,deproj}$ = 1.65 kpc.    The mass at the half light radius is $M_{\rm half} = 6.7\pm 2.1 \times 10^{7} M_{\odot}$; with a V-band luminosity of 9.4 $\times 10^6 L_{\odot}$ from K10 we find a mass to light ratio of $M/L_{V}$ = 7.1 at the half light radius.  At the outermost radius of $r$ = 1.7 kpc we find a mass of  $M_{\rm} = 1.1 \times 10^{7} M_{\odot}$ which results in a mass to light ratio of  $M/L_{V}$ = 11.5.  This inferred mass-to-light ratio is slightly above that expected from an old stellar population, suggesting the presence of a dark matter component with a similar mass to the stellar component.  

Dispersion based mass estimates are calculated using the formula in \citet{Wolf2010} which relates the line-of-sight velocity dispersion and the dynamical mass at the half-light radius.  Using this formalism, $M_{1/2} = 3G^{-1}\langle \sigma_v^2 \rangle r_{1/2}$ where $r_{1/2}$ is the three dimensional deprojected half-light radius and $\langle \sigma_v^2 \rangle$ is the raw velocity dispersion.  Using the same value for the deprojected half light radius in the previous section, where $r_{1/2,deprojected}$ = 1.6 kpc we calculated a dispersion mass of $M_{1/2}$ = 6.5$\pm$1.9 $\times 10^7 M_{\odot}$.  This is approximately the same value as the rotationally derived mass estimate.  The calculated mass to light ratio of $M/L_{V}$ =  6 is similar to the classical MW dSphs Fornax, Leo I and Sculptor, which have  $M/L_{V}$ = 5,  $M/L_{V}$ = 10, and $M/L_{V}$ = 6 respectively (K10).  Again, we stress that these values are based on simple estimators with an assumption of dynamical stability.

\section{Discussion and Conclusions}

We have presented kinematical profiles of the M31 dSph, And II using Keck/DEIMOS spectroscopy of 531 individual stars.  These profiles represent the largest and most spatially complete spectroscopic sample of any M31 dSph and allows us to determine in detail the kinematical behavior of And II.  The derived velocity profile along the major axis show that it possesses no rotation and is dispersion dominated.  Looking near the minor axis, we find that its kinematics is dominated by coherent rotation which has a maximum rotational velocity of $v_{\rm r,max}$ = 8.6$\pm$1.8 \kms\ at the outermost radius and a flat velocity dispersion with an average dispersion, $\langle \sigma_v \rangle$ = 7.5$\pm$1.1 \kms.  Quantifying the degree of misalignment between the photometric and kinematical major axes we find $\Psi = 67^\circ \pm 12^\circ$, which further points to And II being a prolate rotator.  With only a handful of other dwarfs in the LG showing significant rotation, And II is the only object with minor axis rotation, making it a unique system.  

The presence of coherent rotation in And II makes it an interesting object, however this observed rotation may not be intrinsic to the system if it has had a recent encounter with M31.  Tidal interactions between a satellite and its host system can disturb the stellar kinematics of the dwarf system and create unbound tidal tails.  In the M31 system, NGC 205 is an example of a dwarf which is currently tidally interacting with the M31 halo.  The isophotes of this galaxy show strong twists which point toward M31 along with a rotation curve which turns over in the outer radius due to unbound stars \citep{Geha2006a}.  \citet{Munoz2006} showed that for the Carina dSph, unbound stars which were mistakenly assumed to be bound, created an artificial rise in the velocity profile.  When observations at larger radii were included, this velocity gradient disappears and the stellar velocity profile shows a decline in the outer radii.  Thus, in order to properly characterize the kinematical properties of these objects, datasets which span a large radial extent are needed.    Additionally, depending on the current orbit of the system, the tidal tails may be aligned directly with the line of sight \citep{Klimentowski2009}.  However, the densities of these tidal tails are low and thus, the probability of observing these tidally unbound stars, are even lower.  Using And II's current projected distance of 185 kpc and the previously derived dynamical mass, we show that our data, which extends out to a maximum 2.5 $r_{\rm eff}$, is well within the instantaneous tidal radius of $~ 7$ kpc, or $~$7 $r_{\rm eff}$.  Coupling this with the fairly regular rotation that extends out to about 2.5$r_{\rm eff}$ and the lack of abrupt turnovers in the velocity profile, we do not expect the observed minor axis rotation in And II to be tidally induced.  

Determining the dynamical stability of And II is complicated by its peculiar kinematics, but normal metallicity behavior.  The observed radial metallicity gradient, which has also been observed in many other LG dwarfs, is a likely product of increased chemical enrichment in the central regions due to higher central stellar densities.  In studying the chemo-dynamics of the system, we find no significant difference between the velocity and dispersion profiles of the metal-rich and metal-poor component, thus suggesting that And II is comprised of one single stellar component. Along with the rotation being fairly regular, these all point to a system which is dynamically stable.  However, whether a galaxy can naturally possess and maintain such strong minor axis rotation is not well constrained.  \citet{Binney1978} showed that for an ellipticity of $\epsilon$ = 0.1, the distribution of expected $v/\sigma$ values peaks near $v/\sigma$ = 0.3 and quickly drops to near zero at $v/\sigma \sim$1.   With the scarcity of observed systems which have fast, minor axis rotation like And II, the question of its dynamical state can only be answered with more detailed modeling.

While And II is currently an observationally scarce object, work by \citet{Mayer2001a,Mayer2001b, Mayer2006, Lokas2010} exploring the evolution of dwarf irregulars (dIrrs) into dSphs indicate that its existence is not unexpected.  In this tidal stirring scenario, a gas-rich dwarf infalls into a parent halo on an eccentric orbit with a typical perihelion:aphelion ratio of 1:5.  Ram pressure stripping due to the interaction between the host's hot, gaseous halo and the cold, gas disk of the dwarf causes the dwarf to lose most of its gas.  As the dwarf approaches perihelion, the tidal interactions with the host halo increases dramatically and the system undergoes severe tidal disturbance.  These tidal shocks simultaneously strip stars outside the tidal radius while also heating up the dwarf.  With enough tidal heating, a bar instability can be induced.  The bar decays once the prolate stellar orbits which feed it gain enough anisotropy.  This prolate shape can last for up to a Gyr, depending on the orbital parameters of the dwarf relative to its parent halo.  Thus, a prolate rotator, in this scenario, is a transition step between rotationally dominated dIrrs and pressure-supported dSphs.  Whether And II represents this class of transition objects or is just an isolated case of a dwarf galaxy with odd kinematics remains an open question.  Future spectroscopic work on the stellar kinematics of more LG dwarfs will help identify where And II belongs in the dwarf galaxy evolution sequence.

In summary, we studied the stellar kinematics of the M31 dSph And II and found the surprising result that it is a rotationally dominated, prolate system.  Only a handful of dwarfs in the LG have been shown to possess rotation; And II is the first to show minor axis rotation.  Very few massive galaxies have minor axis rotation dominate the dynamics of the system, And II is by far the lowest luminosity galaxy to show this. Whether the system evolved on its own to this state or has been guided by external factors is an open question because while the dynamics of the system is unusual, the chemistry show a fairly regular dwarf that is similar to other LG dwarfs.  It may either be an extreme outlier in the context of dwarf galaxy evolution or is a galaxy in the process of transitioning from a gas-rich, rotationally supported dwarf irregular to a gas-poor, dispersion supported dwarf spheroidal. 

\section{Acknowledgements}

NH acknowledges support from the Connecticut Space Grant Consortium.  M.G. acknowledges support from NSF grant AST-0908752 and the Alfred P. Sloan Foundation.  R.R.M. acknowledges support from the GEMINI-CONICYT Fund, allocated to the project N0 32080010, from CONICYT through project BASAL PFB-06 and the Fondo Nacional de Investigaci\'on Cient\'õÞca y Tecnol\'ogica (Fondecyt project N0 1120013).  PG acknowledges support from NSF grant AST-1010039 to UCSC.  K.M.G. acknowledges support provided by NASA through Hubble Fellowship grant 51273.01 awarded by the Space Telescope Science Institute, which is operated by the Association of Universities for Research in Astronomy, Inc., for NASA, under contract NAS 5-26555.   EJT acknowledges that the support for this work was provided by NASA through Hubble Fellowship grant 51316.01 awarded by the Space Telescope Science Institute, which is operated by the Association of Universities for Research in Astronomy, Inc., for NASA, under contract NAS 5-26555.

\begin{deluxetable*}{lccrrccc}[!b]
\tabletypesize{\scriptsize}
\tablecaption{Keck/DEIMOS Multi-Slitmask Observing Parameters}
\tablewidth{0pt}
\tablehead{
\colhead{Mask} &
\colhead{Date Observed}&
\colhead{$\alpha$ (J2000)} &
\colhead{$\delta$ (J2000)} &
\colhead{PA} &
\colhead{$t_{\rm exp}$} &
\colhead{\# of slits} &
\colhead{\% useful} \\
\colhead{Name}&
\colhead{}&
\colhead{(h$\,$:$\,$m$\,$:$\,$s)} &
\colhead{($^\circ\,$:$\,'\,$:$\,''$)} &
\colhead{(deg)} &
\colhead{(sec)} &
\colhead{}&
\colhead{spectra}
}
\startdata
d2\_1  &  Sept 6, 2005 & 01:17:07.5  & +33:29:25.1 & $-$90 & 3600 & 121 & 63$\%$ \\
d2\_2  & Sept 6, 2005  & 01:16:43.3  & +33:34:25.8 & 0 & 3600 &141 & 52$\%$ \\
d2\_3  & Oct 11, 2007 & 01:17:07.7 & +33:22:14.8 & $-$90 & 3600 & 130 & 15$\%$ \\
d2\_4  & Oct 11, 2007 & 01:16:33.8 & +33:19:29.7 & $-$180 & 3600 & 146 & 45$\%$ \\
d2\_5  & Oct 11, 2007 & 01:15:43.8 & +33:24:26.8 & +90 & 3600 & 133 & 41$\%$ \\
d2\_6  & Oct 12, 2007 & 01:16:17.6 & +33:17:21.3 & +180 & 3600 & 144 & 51$\%$ \\
d2\_7 & Oct 12, 2007  & 01:15:43.5 & +33:27:19.1 & +90 & 3600 & 125 & 51$\%$ \\
d2\_8 & Oct 12, 2007  & 01:16:15.9 & +33:34:12.6 & 0 & 3600 & 137 & 62$\%$ \\
d2\_9 & Oct 11, 2007  & 01:17:03.1 & +33:25:47.8 & 0 & 3100 & 140 & 63$\%$ \\
d2\_10 & Oct 12, 2007  & 01:15:41.6 &+33:25:53.8 & 0 & 3700 & 135 & 60$\%$ \\
d2\_11 & Aug 24, 2009 & 01:16:19.4 & +33:27:21.3 & $-$45 & 3600 & 174 & 40$\%$ \\
d2\_12 & Nov 16, 2009 & 01:16:28.8 & +33:24:14.4 & 0 & 4800 & 117 & 84$\%$ 
\enddata
\tablecomments{The observation date, right ascension, declination,
  position angle and total exposure time for each Keck/DEIMOS
  slitmask.  The final two columns refer to the total number of
  slitlets on each mask and the percentage of those slitlets for which
  a redshift was measured.} \label{table_mask}
\end{deluxetable*} 

\begin{deluxetable*}{cccccc}[!h]
\tabletypesize{\scriptsize}
\tablecaption{Isophotal Parameters}
\tablewidth{0pt}
\tablehead{
\colhead{Radius} &
\colhead{Number Density} &
\colhead{Ellipticity} &
\colhead{Ellipticity} &
\colhead{PA}&
\colhead{PA}\\
\colhead{(arcmin)}&
\colhead{$\#$/arcmin$^2$}&
\colhead{(IRAF)} &
\colhead{(Circular Annulus)}&
\colhead{(IRAF)} &
\colhead{(Circular Annulus)} 
}
\startdata
1 & 436$\pm$20 & 0.17$\pm$0.04 & 0.20$\pm$0.04 & 38$\pm$7 & 17$\pm$3  \\
2 & 307$\pm$17 & $..$ & $..$ & $..$ & $..$  \\
3 & 181$\pm$13 & $..$ & $..$ & $..$ & $..$  \\
4 & 126$\pm$11 & 0.09$\pm$0.05& 0.24$\pm$0.06 & 49$\pm$16 & 73$\pm$4 \\
5 & 97$\pm$10 & 0.18$\pm$0.03 & 0.22$\pm$0.03 & 33$\pm$6 &18$\pm$6\\
6 & 77$\pm$9 & 0.23$\pm$0.04& 0.18$\pm$0.02 & 48$\pm$5 & 24$\pm$4\\
7 & 65$\pm$8 & 0.26$\pm$0.03& 0.24$\pm$0.01 & 52$\pm$4 & 27$\pm$6\\
8 & 48$\pm$7 & 0.24$\pm$0.02& 0.22$\pm$0.03 & 57$\pm$3 & 38$\pm$5\\
9 & 40$\pm$6 & 0.23$\pm$0.02 & 0.10$\pm$0.02 & 62$\pm$3 & 26$\pm$9 \\
10 & 30$\pm$5 & 0.11$\pm$0.02& 0.08$\pm$0.03 & 65$\pm$5 & 74$\pm$6 \\
11 & 21$\pm$5 & 0.07$\pm$0.02 & 0.13$\pm$0.02 & 65$\pm$8 & 80$\pm$5\\
12 & 14$\pm$4 & 0.04$\pm$0.02 & 0.10$\pm$0.03 & 58$\pm$14 & 43$\pm$8 \\
13 & 10$\pm$3 & 0.07$\pm$0.01 & 0.10$\pm$ 0.02 & 26$\pm$6 &46$\pm$8\\

\enddata
\tablecomments{Derived isophotal parameters for And II using archival Subaru Suprime-Cam data.  The number density, as well as position angle and ellipticity are presented for each radius, in steps of one arcminute.  Ellipticity and position angle derivations from both the circular annulus method and IRAF \text{ellipse} are presented.  Derivations of ellipticity and position angle were omitted for radii 2$'$ and 3$'$ because of extensive masking of the regions due to saturated stars.}\label{isophotal_parameters}
\end{deluxetable*} 

\begin{deluxetable*}{lllcccc}[!h]
\tabletypesize{\scriptsize}
\tablecaption{Kinematical Major Axis Profile}
\tablewidth{0pt}
\tablehead{
\colhead{x} &
\colhead{$\alpha$ (J2000)} &
\colhead{$\delta$ (J2000)} &
\colhead{V}&
\colhead{V$_{\rm err}$} &
\colhead{$\sigma$} &
\colhead{$\sigma_{\rm err}$} \\
\colhead{arcmin}&
\colhead{(h$\,$:$\,$m$\,$:$\,$s)} &
\colhead{($^\circ\,$:$\,'\,$:$\,''$)} &
\colhead{(\kms)}&
\colhead{(\kms)} &
\colhead{(\kms)} &
\colhead{(\kms)}
}
\startdata

$-$8.8 & 1:16:01.6 & 33:19:16.0 & 1.5 & 1.6 & 7.7 & 1.4 \\
$-$7.1 & 1:16:06.5 & 33:20:32.3 & 1.9 & 1.9 & 10.6 & 1.5 \\
$-$5.3 & 1:16:11.4 & 33:21:48.5 & $-$0.4 & 1.9 & 11.6 & 1.5 \\
$-$3.5 & 1:16:16.3 & 33:23:04.7 & 1.4 & 2.2 & 12.7 & 1.6 \\
$-$1.8 & 1:16:21.2 & 33:24:20.9 & 1.2 & 1.8 & 11.0 & 1.4 \\
00.0 & 1:16:26.1 & 33:25:37.2 & 2.0 & 1.3 & 8.7 & 1.1 \\
$+$1.8 & 1:16:31.0 & 33:26:53.4 & $-$0.1 & 1.9 & 11.3 & 1.5 \\
$+$3.5 & 1:16:35.9 & 33:28:09.7 & $-$4.3 & 1.7 & 11.8 & 1.3 \\
$+$5.3 & 1:16:40.9 & 33:29:25.9 & $-$0.8 & 1.2 & 6.5 & 1.1 \\
$+$7.1 & 1:16:45.8 & 33:30:42.1 & $-$0.3 & 1.3 & 7.4 & 1.1 \\
$+$8.8 & 1:16:50.7 & 33:31:58.4 & $-$0.9 & 1.2 & 6.7 & 1.0

\enddata
\tablecomments{The major axis velocity profile of And II with velocity (V) and velocity dispersion ($\sigma$) as a function of projected distance, x, along the major-axis from the galaxy center.  Positive x values correspond to the Northwest semi-major axis of the galaxy, while negative x values correspond to the Southwestern semi-major axis.  }\label{major_axis_data}
\end{deluxetable*} 
 
\begin{deluxetable*}{lllcccc} [!h]
\tabletypesize{\scriptsize}
\tablecaption{Kinematical Minor Axis Profile}
\tablewidth{0pt}
\tablehead{
\colhead{y} &
\colhead{$\alpha$ (J2000)} &
\colhead{$\delta$ (J2000)} &
\colhead{V}&
\colhead{V$_{\rm err}$} &
\colhead{$\sigma$} &
\colhead{$\sigma_{\rm err}$} \\
\colhead{arcmin}&
\colhead{(h$\,$:$\,$m$\,$:$\,$s)} &
\colhead{($^\circ\,$:$\,'\,$:$\,''$)} &
\colhead{(\kms)}&
\colhead{(\kms)} &
\colhead{(\kms)} &
\colhead{(\kms)}
}
\startdata

$-$8.8 & 1:16:51.5 & 33:19:29.1 & 7.5 & 1.8 & 4.8 & 1.8\\
$-$7.1 & 1:16:46.5 & 33:20:42.7 & 11.6 & 1.7 & 5.8 & 1.4\\
$-$5.3 & 1:16:41.4 & 33:21:56.3 & 7.6 & 1.5 & 8.2 & 1.1\\
$-$3.5 & 1:16:36.3 & 33:23:09.9 & 4.0 & 1.4 & 9.9 & 1.1\\
$-$1.8 & 1:16:31.2 & 33:24:23.6 & 3.5 & 0.9 & 5.9 & 0.8\\
00.0 &   1:16:26.1 & 33:25:37.2 & 1.4 & 1.1 & 8.1 & 0.9\\
$+$1.8 & 1:16:21.1 & 33:26:50.8 & -1.9 & 1.3 & 8.8 & 1.1\\
$+$3.5 & 1:16:15.9 & 33:28:04.4 & -6.6 & 1.3 & 7.5 & 1.1\\
$+$5.3 & 1:16:10.9 & 33:29:18.0 & -4.3 & 1.9 & 10.8 & 1.5\\
$+$7.1 & 1:16:05.8 & 33:30:31.7 & -8.8 & 1.6 & 8.4 & 1.3\\
$+$8.8 & 1:16:0.70 & 33:31:45.3 & -9.6 & 1.8 & 8.5 & 1.6\\
\enddata
\tablecomments{The minor axis velocity profile of And II with velocity (V) and velocity dispersion ($\sigma$) as a function of projected distance, y, along the minor-axis from the galaxy center.  Positive y values correspond to the Northwest semi-minor axis while negative y values correspond to the Southeastern semi-minor axis. }\label{minor_axis_data}
\end{deluxetable*}

\end{document}